\documentclass[twocolumn,10pt,aps,prb]{revtex4}   

\usepackage{amsmath}    % need for subequations
\usepackage{graphicx}   % for figures

\begin{document}

\title{Memristor for introductory physics}
\author{Frank Y.~Wang} \affiliation{Mathematics Department, LaGuardia
  Community College, The City University of New York, Long Island
  City, New York 11101} \email{fwang@lagcc.cuny.edu} %optional

\maketitle

%\section{Introduction}

Students learn from physics textbooks that there are three fundamental
two-terminal circuit elements: resistors, capacitors and inductors.
In 2008 May, scientists at Hewlett-Packard Laboratories published a
paper in \emph{Nature} announcing the invention of the fourth
element---memristor.\cite{fourth,nature} In that paper, Strukov et al
presented a model to illustrate their invention.  Their model provides
a simple explanation for several puzzling phenomena in nanodevices,
yet it involves no more mathematics than basic algebra and simple
integration and differentiation, and is comprehensible for college
students with knowledge of first-year physics and calculus.  In this
note, we guide students to derive the analytical solution to the
equations for the memristor model described in that paper, and use it
to calculate the current responding to a sinusoidal voltage.  Through
this exercise, students will realize the nonlinear effects in electric
circuit, which is ubiquitous in nanoscale electronics.

The definition of electric current $i$ is the time derivative of
electric charge $q$, and according to Faraday's law voltage $v$ is the
time derivative of magnetic flux $\varphi$.  In 1971 Leon Chua noted
that there ought to be six mathematical relations connecting pairs of
the four fundamental circuit variables, $i$, $v$, $q$ and
$\varphi$.\cite{chua71} In addition to the definition of electric
current and Faraday's law, three circuit elements connect pairs of the
four circuit variables: resistance $R$ is the rate of change of
voltage with current ($R=\frac{d v}{di}$), capacitance $C$ is the rate
of change of charge with voltage ($C=\frac{dq}{dv}$), and inductance
$L$ is the rate of change of magnetic flux with current ($L=\frac{d
  \varphi}{d i}$).  From symmetry arguments, Chua reasoned that there
should be a fourth fundamental element, which he called a memristor
(short for memory resistor) $M$, for a functional relation between
charge and magnetic flux, $M=\frac{d \varphi}{dq}$.

\begin{figure}[h]
\begin{center}
\includegraphics[scale=1]{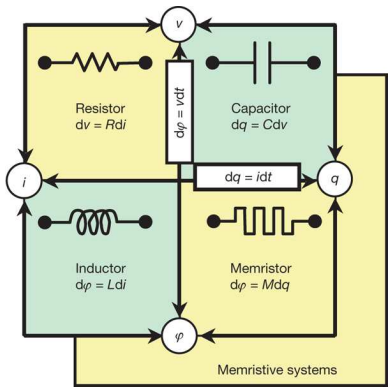}
\caption{Figure adapted from \emph{Nature}.}
\end{center}
\end{figure}

In the case of linear elements, in which $M$ is a constant,
memristance is identical to resistance.  However, if $M$ is itself a
function of $q$, yielding a nonlinear circuit element, then no
combination of resistive, capacitive and inductive circuit elements
can duplicate the properties of a memristor.  Almost 40 years after
Chua's proposal, memristor was found by scientists at HP Labs led
by R. Stanley Williams.\cite{fourth,nature} They suspected that
memristive phenomenon has been hidden for so long because those
interested were searching in the wrong places.  Although the definition
of memristor involves magnetic flux, magnetic field does not play an
explicit role in the mechanism of memristance.  As stated in their
paper, \emph{the mathematics simply require there to be a nonlinear
  relationship between the integrals of the current and voltage}
[italics added]. The explicit relationship between such integrals
based on the model by Strukov et all is detailed below.

After the original proposal of the memristor, Chua and Kang
generalized the concept to a broader class of systems, called
memristive systems,\cite{chua76} defined as
\begin{equation}\label{eq:gohm}
v = \mathcal{R}(x) \, i
\end{equation}
\begin{equation}\label{eq:memristive}
\frac{dx}{dt} = f(x,i)
\end{equation}
Eq.~(\ref{eq:gohm}) looks like Ohm's law, but the generalized
resistance $\mathcal{R}(x)$ depends upon the internal state $x$ of the
device.  The time derivative of the internal state is a function of
$x$ and $i$.  Strukov et al were the first to conceive a physical
model in which $x$ is simply proportional to charge $q$.  They
designed a device with $\mathcal{R}$ that can switch reversibly
between a less conductive OFF state and a more conductive ON state,
depending on $x$
\begin{equation}
\mathcal{R}(x) = x(t) R_{\text{on}} + [1-x(t)] R_{\text{off}}
\end{equation}
The internal state $x(t)$ is restricted in the interval $[0,1]$; when
$x(t) = 0$, $\mathcal{R} = R_{\text{off}}$, and when $x(t)=1$,
$\mathcal{R} = R_{\text{on}}$.  The time derivative of $x(t)$ is made
to be proportional to current
\begin{equation}\label{eq:utimederivative}
\frac{dx(t)}{dt} = \frac{R_{\text{on}}}{\beta} i(t)
\end{equation}
in which the parameter $\beta$ has a dimension of magnetic flux (in SI
units, magnetic flux is measured in V s, or Wb).  The HP scientists
produced a semiconductor device to satisfy this condition, to be
discussed shortly.  Eq. (\ref{eq:gohm}) is rewritten as
\begin{equation}\label{eq:rde}
v(t) = \{ x(t) R_{\text{on}} + [1-x(t)] R_{\text{off}} \} i(t)
\end{equation}
Eliminating $i(t)$ from Eq.~(\ref{eq:rde}) using
Eq.~(\ref{eq:utimederivative}), we can write the memristive system as
a first-order differential equation of $x(t)$.  Defining the
resistance ratio $r=R_{\text{off}}/R_{\text{on}}$, the equation
becomes
\begin{equation}
v(t) = \beta \{ x(t) + r [1-x(t)] \}  \frac{d x(t)}{dt}
\end{equation}
Using $\varphi=\int v dt$, and the property $x \frac{dx}{dt} =
\frac{1}{2} (\frac{d}{dt} x^{2})$, it is easy to integrate both sides
\begin{equation}\label{eq:vintandiint}
\varphi =
\beta [ - \frac{r-1}{2} x^{2}+ r x + c]
\end{equation}
where $c$ is a constant of integration determined by the initial value
of $x$.  This constant contains the past events of the input current
$i$ and plays an important role in a nonlinear system.  Since $x$ is
proportional to charge, Eq.~(\ref{eq:vintandiint}) indicates that the
magnetic flux is a quadratic function of the charge, and the nonlinear
relationship between the integrals of the current and voltage for a
memristor is achieved.

The memristor invented at HP Labs is made of a thin film of
titanium dioxide of thickness $D$ sandwiched between two platinum
contacts.  The film has a region with a high concentration of dopants
having low resistance $R_{\text{on}}$, and the remainder has a low
dopant concentration and much higher resistance $R_{\text{off}}$.  The
application of an external bias $v(t)$ across the device will move the
boundary between the two regions by causing the charged dopants to
drift.  Let $w(t)$ be the coordinate of the boundary between a less
conductive OFF state and a more conductive ON state, and $\mu_{V}$ the
average ion mobility (measured in m$^{2}$~s$^{-1}$~V$^{-1}$ in SI
units), Eq.~(\ref{eq:utimederivative}) written in physical parameters
is
\begin{equation}\label{eq:dwdt}
\frac{1}{D} \frac{d w(t)}{dt} = \mu_{V} \frac{R_{\text{on}}}{D^{2}} i(t)
\end{equation}  
which leads to
\begin{equation}
\frac{w(t)}{D} = \mu_{V} \frac{R_{\text{on}}}{D^{2}} q(t)
\end{equation}    
By comparing Eq.~(\ref{eq:utimederivative}) and Eq.~(\ref{eq:dwdt}) we
identify $x=w/D$ and $\beta=D^{2}/\mu_{V}$.  Recall that $\beta$ has a
dimension of magnetic flux, thus magnetic field is not directly
involved in this memristor.  Eq.~(\ref{eq:vintandiint}), setting $c$
to zero, becomes
\begin{equation}
\varphi = -\frac{R_{\text{on}} \mu_{V}}{2 D^{2}} 
\left(\frac{R_{\text{off}}}{R_{\text{on}}} -1 \right) q^{2} +
R_{\text{off}} q 
\end{equation}
For $R_{\text{off}}\gg R_{\text{on}}$, the memristance of this device
is obtained
\begin{equation}
M(q) \equiv \frac{d \varphi}{dq} = 
R_{\text{off}} \left(1-\frac{\mu_{V} R_{\text{on}}}{D^{2}} q \right)
\end{equation}
The $q$-dependent term is due to the nonlinear relationship between
$\varphi$ and $q$.  For the first time, scientists wrote down a
formula for the memristance of a device in terms of material and
geometrical properties of the device.  (An analogy is that for
parallel plates, capacitance is expressed as $C= \epsilon A/d^{2}$,
where $\epsilon$, $A$ and $d$ are permittivity, plate area and plate
separation, respectively.  Similarly resistance and inductance of
devices can be expressed in terms of material and geometrical
properties.)  Because the coefficient of $q$ is proportional to
$1/D^{2}$, memristive behavior is increasingly relevant as electronic
devices shrink to a width of a few nanometers.  As semiconductor film
of thickness $D$ reduces from micrometer scale to nanometer scale, the
nonlinear effect increases by a factor of 1,000,000.

\begin{figure}[h]
\begin{center}
\includegraphics[scale=0.45]{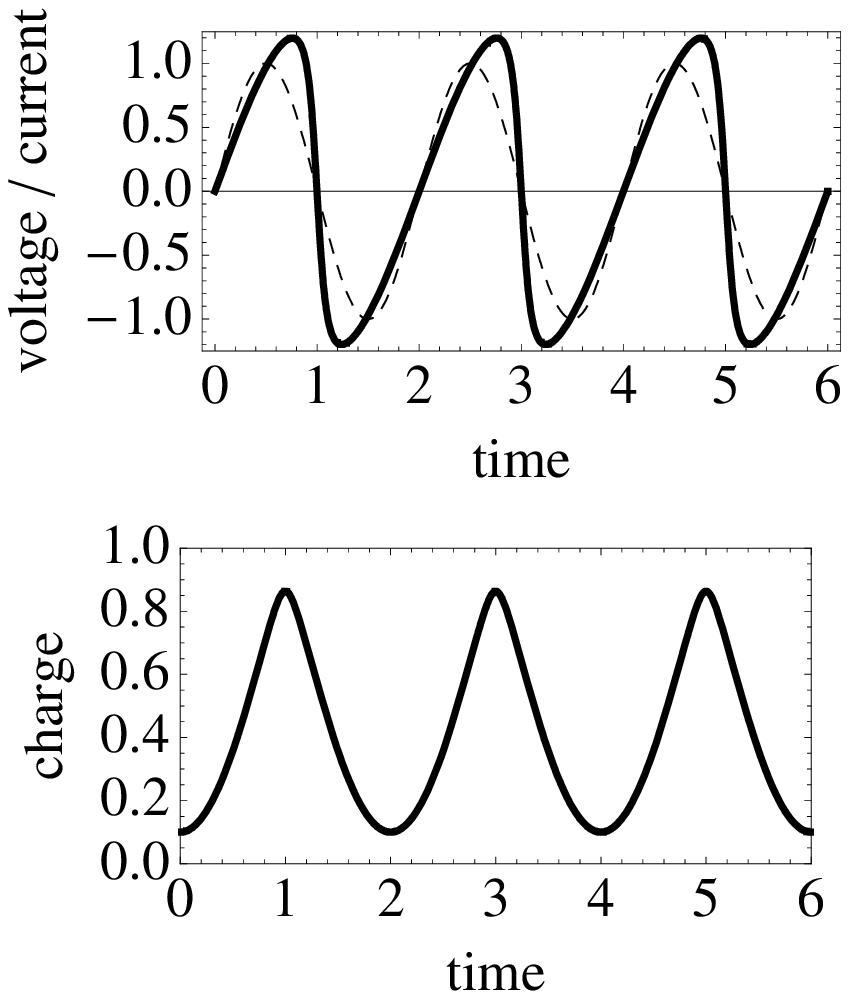}
\includegraphics[scale=0.4]{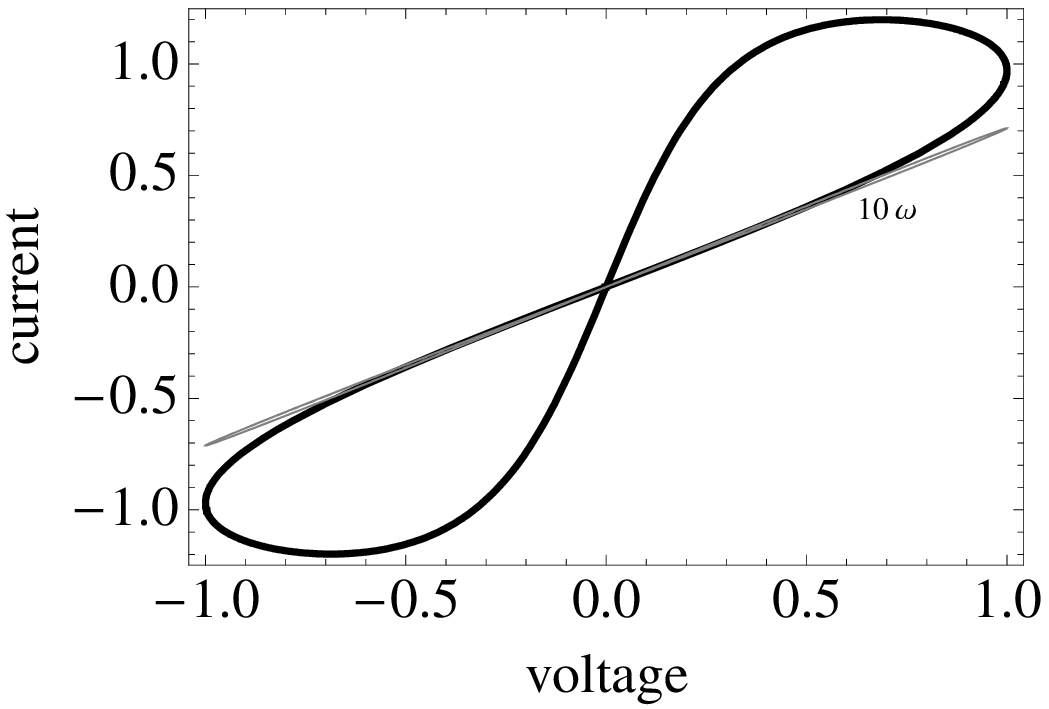}
\includegraphics[scale=0.4]{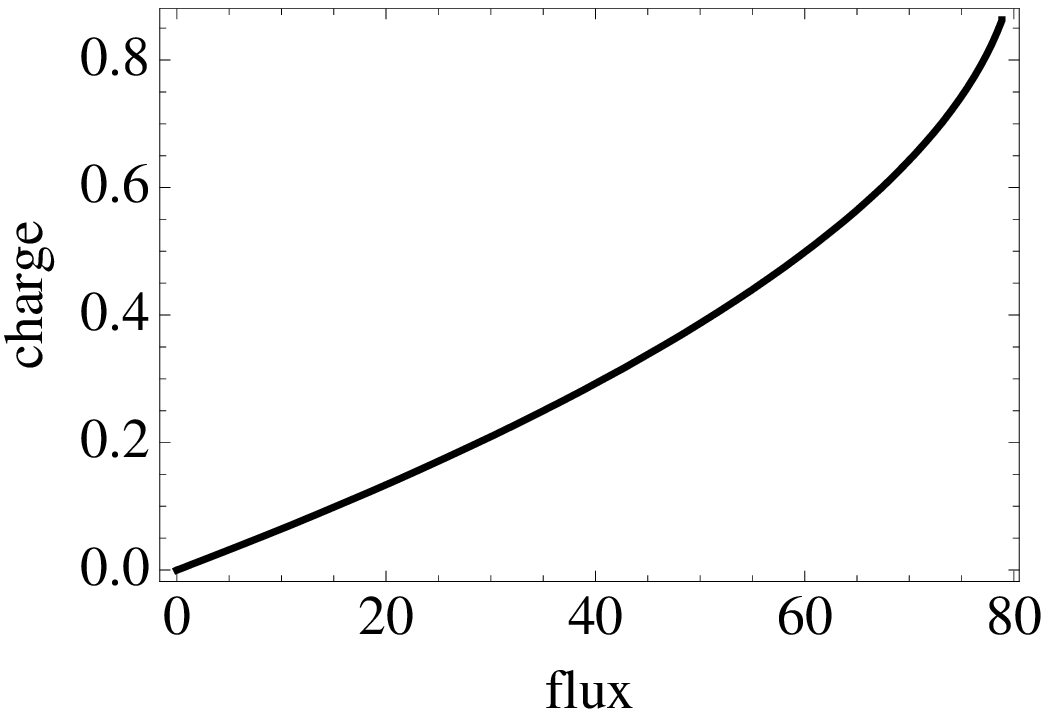}
\caption{\label{fig:2b} The charge and current of a memristor
  responding to the applied voltage $v_{0} \sin (\omega t)$ (dashed
  line in the $v$--$t$ plot) based on Eqs.~(\ref{eq:charge}) and
  (\ref{eq:iresponded}) are shown; the $q$--$\varphi$ plot is made
  using Eq.~(\ref{eq:vintandiint}).  In these plots, voltage, current,
  time, magnetic flux and charge are expressed in units of $v_{0}$,
  $v_{0}/R_{\text{on}}$, $t_{0} \equiv \beta/v{0}$, $\beta$ and $i_{0}
  t_{0}$, respectively.  The resistance ratio is
  $r=R_{\text{off}}/R_{\text{on}}=160$; the dimensionless combination
  of $v_{0}/(\beta \omega)$ is set to be $100/\pi$, and $\beta$
  (physically $D^{2}/\mu_{V}$ in which $D$ is the film thickness and
  $\mu_{V}$ the mobility) is chosen to be $10^{-2}$ Wb.  The $i$--$v$
  plot is produced using the parametric plot procedure in graphing
  software.  In the $i$--$v$ plot, a high-frequency ($10 \omega$)
  response is shown, which appears almost as a straight line.}
\end{center}
\end{figure}

\begin{figure}[t]
\begin{center}
\includegraphics[scale=0.45]{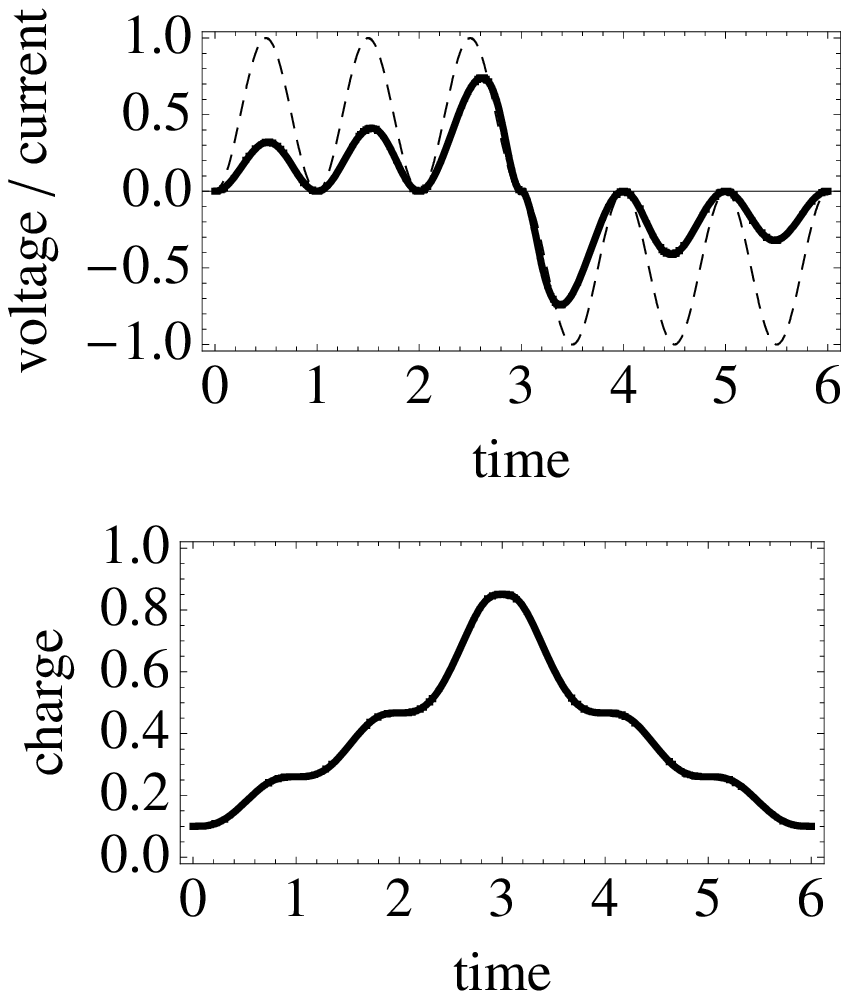}
\includegraphics[scale=0.4]{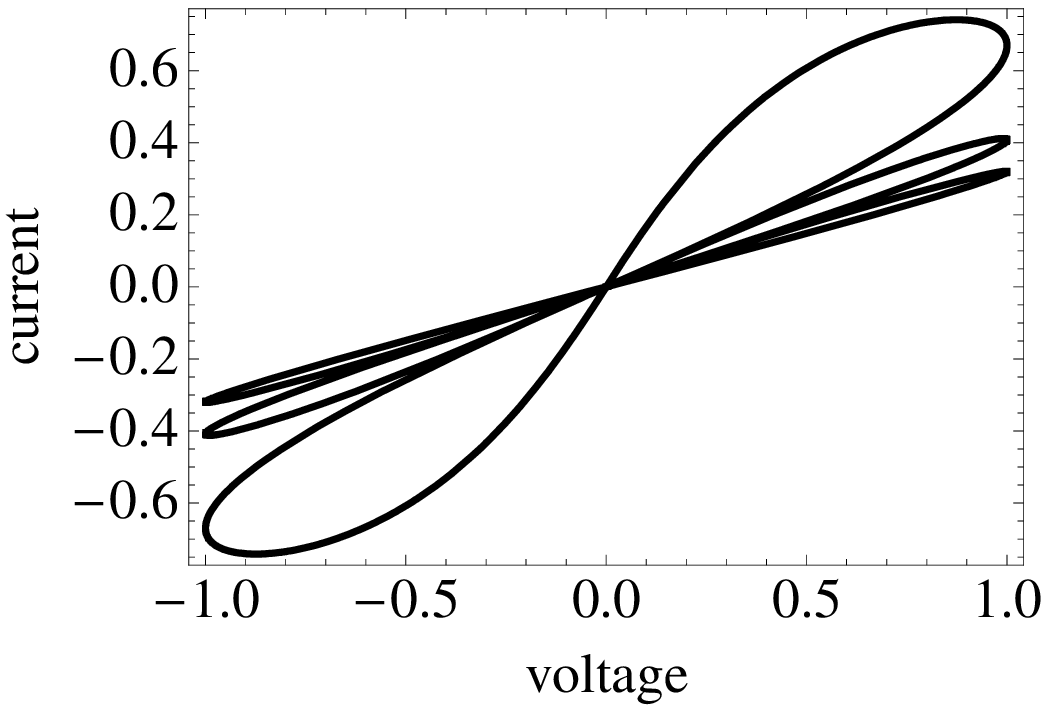}
\includegraphics[scale=0.4]{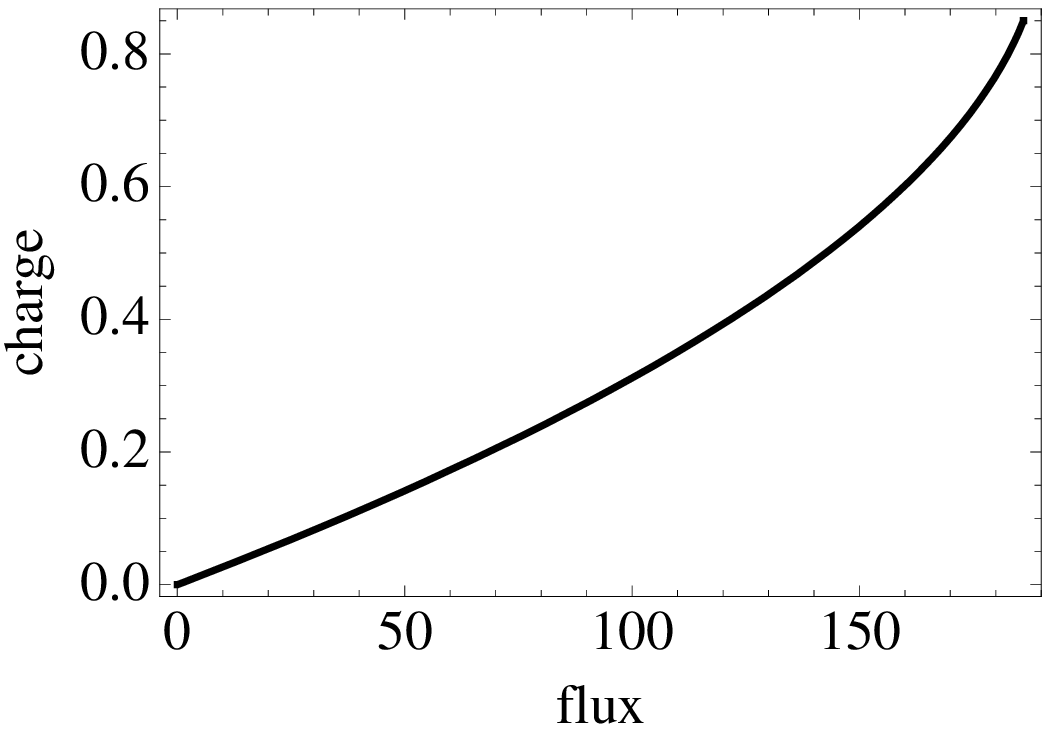}
\caption{\label{fig:2c} This figure is similar to Figure~\ref{fig:2b}:
  the applied voltage is $\pm v_{0} \sin^{2} (\omega t)$ (dashed line
  in the $v$--$t$ plot), and resistance ratio
  $R_{\text{off}}/R_{\text{on}}=380$.  Notice that while $i$--$v$ plot
  in this example appears to be complex, the pinched loops feature
  manifests itself vividly.  Also the $q$--$\varphi$ plot is simple:
  magnetic flux is a single-valued function of charge, see
  Eq.~(\ref{eq:vintandiint}).}
\end{center}
\end{figure}

As soon as researchers discovered the device, it is natural to measure
the current responding to the applied voltage.  If the applied voltage
in Eq.~(\ref{eq:rde}) is
\begin{equation}\label{eq:vapplied}
v(t) = v_{0} \sin (\omega t)
\end{equation}
then replacing $\varphi=\int v(t) dt$ with $-\frac{v_{0}}{
  \omega} \cos(\omega t)$ in Eq.~(\ref{eq:vintandiint}) and using the
quadratic formula to solve for $x(t)$, we have
\begin{equation}\label{eq:charge}
x(t) = \frac{r - \sqrt{r^{2} + 2 (r-1)
\left[\frac{v_{0}}{\beta \omega} \cos(\omega t)+c\right]}}{r-1}
\end{equation}
(We retain only the sign ``$-$'' in ``$\pm$'' so that the solution is
in the interval $[0,1]$.)  The current $i(t)$ is proportional to the
time derivative of $x(t)$, which is
\begin{equation}\label{eq:iresponded}
\frac{d}{dt} x(t) = \frac{v_{0}}{\beta}
\frac{\sin (\omega t)}{
\sqrt{r^{2} + 2 (r-1)
\left[\frac{v_{0}}{\beta \omega} \cos(\omega t)+c\right]}}
\left[ =\frac{R_{\text{on}}}{\beta} i(t) \right]
\end{equation}
Our calculation shows that the current is the product of $v(t)$ and a
function of $\cos (\omega t)$.  Important properties of a memristor
can be understood from this expression.  The current $i(t)$ is zero
whenever the the applied voltage $v(t)$ is zero, regardless of the
state $x(t)$.  Because the $i$--$v$ curve must pass through the
origin, the $i$--$v$ Lissajous figure is double-loop curve---the most
salient feature of the memristor.  At high frequencies, $\cos(\omega
t)/\omega$ under the radical sign becomes small, so the nonlinear
contribution is suppressed.  As a result, the $i$--$v$ hysteresis
collapses to a nearly straight line.  These features are seen in
Figure~\ref{fig:2b}.  The memory effect of a memristor is attributed
to the constant of integration $c$ first appeared in
Eq.~(\ref{eq:vintandiint}).  This constant cannot be removed by
subtracting a constant from $x(t)$, and remains after differentiating
$x(t)$ with time.  Therefore, memristor response depends on history.
As another example, with the applied voltage $v(t) = \pm v_{0}
\sin^{2} (\omega t)$ shown in Figure~\ref{fig:2c}, the solution is
obtained by replacing $\varphi = \pm \int v_{0} \sin^{2} (\omega t) dt
= \pm v_{0} [t/2 - \cos(\omega t) \sin(\omega t)/(2 \omega)]$.  We
again see the ``pinched loops'' in the $i$--$v$ plot.  Strukov et al
reported in their paper that $i$--$v$ plots shown Figures~\ref{fig:2b}
and \ref{fig:2c} are regularly observed in their studies of titanium
dioxide devices.

The most important lesson for students to learn from this note is that
the familiar Ohm's law $v=iR$ is merely an approximation which is
inadequate for a nonlinear circuit.  Suppose that a scientist, who did
not know the nature of the titanium dioxide device invented by the HP
scientists, attempts to characterize this device on the $i$--$v$
plane.  If one treats the system as resistive, then based on
Eqs.~(\ref{eq:vapplied}) and (\ref{eq:iresponded}), the resistance
would be
\begin{equation}
R = R_{\text{on}}  \sqrt{r^{2} + 2 (r-1)
\left[\frac{v_{0}}{\beta \omega} \cos(\omega t)+c\right]}
\end{equation} 
which is time-varying and frequency-dependent.  In fact, such an
erroneous interpretation of the $i$--$v$ plot is the source of
numerous confusions of ``current-voltage anomalies'' reported in
unconventional elements.\cite{fourth,nature,chua03}  For a memristor,
the constitutive relation is a curve in the $q$--$\varphi$ plane, as
depicted in Figures~\ref{fig:2b} and \ref{fig:2c}, expressed as a
single-valued function of the charge $q$  
\begin{equation}\label{eq:constitutive}
\varphi = \hat{\varphi}(q)
\end{equation}
[This is a rather awkward notation to indicate that $\varphi$ is a
  function of $q$.]  Differentiating this equation with time results
in
\begin{equation}
\frac{d}{dt} \varphi = v = \frac{d \hat{\varphi}}{d q} \frac{dq}{dt}
=M(q) i
\end{equation}  
In terms of voltage and current, Eq.~(\ref{eq:constitutive}) assumes
the form $v = M(q) i$.  Because $M(q)$ varies with the instantaneous
value of charge, which book keeps the past events of the input
current, it has a memory.  Without recognizing this fact, it is
difficult to use $v=R i$ to explain phenomena in nonlinear circuits
properly.  In retrospect, many ``unusual hysteresis'' in $i$--$v$
plots of various circuits are actually due to memristance.  With the
model of Strukov et al, a wide variety of eccentric and complex
Lissajous figures that have been observed in many thin-film,
two-terminal devices can now be understood as memristive behavior.
Such a concept is exceedingly useful in understanding of hysteretic
current-voltage behavior observed in nanoscale electronic devices, and
should be introduced to students as early as possible.  It is worth
mentioning that in preparing the paper for \emph{Nature}, Dr. Williams
and his colleagues were intentionally writing for curious
undergraduates.\cite{williams} Students are encouraged to read the
original paper for further discussions and potential applications.

\end{document}